\documentclass[sigplan,screen]{cidr-2025}

\usepackage{enumitem}
\usepackage{booktabs}
\usepackage{url}
\usepackage{tikz}
\usepackage{eso-pic}
\usetikzlibrary{arrows.meta,positioning,fit,backgrounds,calc}
\usepackage{hyperref}
\setcopyright{none}
\renewcommand{\footnotetextcopyrightpermission}[1]{}

\newcommand{\dna}[1]{#1}
\newcommand{\gang}[1]{#1}

\begin{document}

\title[Experience Graphs: The Data Foundation for Self-Improving Agents]{Experience Graphs: The Data Foundation for Self-Improving Agents}
\settopmatter{authorsperrow=1}


\author{Gang Liao$^{\dagger}$ \quad
Yujia He \quad
Abdullah Ozturk \quad
Zhouyang Li \quad
Ying Wang\quad
Zhitong Guo \quad \\
Hongsen Qin \quad
Yaobin Qin \quad
Tao Yang \quad
Zewei Jiang \quad
Dianshi Li \quad
Jort Gemmeke \quad \\
Jiangyuan Li \quad
Liyuan Li \quad
Nathan Yan \quad
Masha Basmanova \quad
Uladzimir Pashkevich \quad \\
Matt Steiner \quad
Pedro Pedreira\quad
Rob Fergus \quad
Anirudh Goyal \quad
Carole-Jean Wu \quad
Gaoxiang Liu$^{\dagger}$}
\affiliation{%
  \institution{Meta Platforms}
  \city{}
  \country{}
}
\email{gangliao@meta.com, gaoxiang@meta.com}

\author{Andrew Witten \quad Daniel J. Abadi$^{\dagger}$}
\affiliation{%
  \institution{University of Maryland, College Park}
  \city{}
  \country{}
}
\email{abadi@umd.edu}

\renewcommand{\shortauthors}{Liao, et al.}
\makeatletter
\gdef\@shortauthors{Liao, et al.}
\gdef\@authorsaddresses{}
\def\@setauthorsaddresses{}
\let\footnotetextauthorsaddresses\@gobble
\makeatother
\date{}

\begin{abstract}
The database community has repeatedly advanced the state of the art by
recognizing that \textit{new workloads demand new system architectures}.
We argue that long-horizon agentic tasks---code generation, scientific
discovery, hardware design, security research---are such a workload.
Rather than producing a single answer, these agents \emph{explore}: they
generate artifacts, execute tools, observe failures, branch, repair, and
compare alternatives over hundreds of steps, improving over a baseline as
they go. This search produces a rich, structured object we call an
\emph{experience graph}: executable artifacts, tool outputs, objective
rewards, sibling comparisons, mutable search statistics, and causal
lineage. Yet existing agent frameworks treat this experience as disposable
state---serialized into JSON checkpoints and session logs that cannot be
recovered after a crash, queried across users, searched by similarity, or
materialized into training data without brittle, after-the-fact scraping.
Even the file-based memory that production agents have converged
on---declarative facts, procedural skills, episodic logs---captures what
an agent \emph{knows}, not the reward-bearing experience graph of what its
search \emph{tried}.

We propose \textsc{Trellis}\footnote{A trellis is the structural lattice that plants climb and grow on---it does not grow itself, but supports and guides all growth. That is the role of the data foundation: agents explore, evolve, and improve; the trellis holds the shape.}: a data foundation that treats the
experience graph as first-class, governed, queryable database state rather
than ephemeral logs. The core insight is that \textit{search over
experience graphs is a database access pattern}. Frontier selection is a
query, cross-session reuse is vector-seeded graph retrieval, training-data
extraction is a materialized view, and reconstructing what an agent knew
at any past step is a time-travel query. When the database owns the
experience graph instead of the agent process, agents become
\textit{stateless, serverless compute}, and crash recovery, horizontal
scaling, and a closed-loop training flywheel emerge as architectural
byproducts. We ground the design in KernelEvolve, a production
accelerator-kernel optimizer at Meta, where cross-session reuse reaches a
target speedup roughly 10$\times$ faster at 52\% lower token cost per valid
solution. More broadly, Trellis turns inference-time search from
disposable computation into a durable institutional asset: logs made
databases reliable; experience graphs may make agents cumulative.

\end{abstract}

\maketitle

\begingroup
\renewcommand\thefootnote{$\dagger$}
\footnotetext{Corresponding authors.}
\endgroup

\section{Introduction}
\label{section:intro}

Database systems tailor-designed for workloads outperform generic systems~\cite{stonebraker2005onesize,stonebraker2018cstore}. Transaction
processing, warehousing, streaming, graphs, and key-value serving have each grown into its own research area once its access patterns were named precisely. We expect that
\textit{long-horizon agentic tasks}~\cite{anthropic2025harnesses,zhipu2026glm} are driving the next wave that transforms database design.

These tasks are solved not by one-shot inference but by \textit{iterative
search}: the agent explores, evaluates, fails, repairs, and refines
solutions over tens to hundreds of steps before converging, each attempt
informed by the ones before it. Because every converged solution raises
the baseline for the next, this loop is a concrete form of
\textit{recursive self-improvement} (RSI)~\cite{sviokla2026rsi,anthropic2026rsi}.
Agentic applications are already adopted in real production environments at scale. For example, AlphaEvolve~\cite{novikov2025alphaevolve} uses evolutionary search
to discover algorithms, optimizes TPU circuits, and improves
Spanner~\cite{corbett2013spanner} compaction
heuristics. Anthropic's Project Glasswing~\cite{glasswing2025} chains
multi-step exploits to find vulnerabilities across every major OS and
browser. Meta's KernelEvolve~\cite{kernelevolve2026,kernelevolve2026arxiv,kernelevolve2026blog} uses
Monte Carlo Tree Search (MCTS)~\cite{kocsis2006mcts,browne2012mcts} and evolutionary search algorithms to
generate optimized kernels for AI hardware---
programs that translate machine learning model operations into chip-specific
instructions---for NVIDIA, AMD, Meta Training Inference Accelerators (MTIA)~\cite{coburn2025mtia}, and CPU
targets, delivering over 60\% inference throughput improvement on
production ranking and recommendation workloads~\cite{kernelevolve2026blog}. A growing ecosystem of
frameworks~\cite{jiang2025aide,toledo2025aira,openevolve2025,yamada2025aiscientist,carlini2026compiler,pepe2026agentic}
has emerged that drives ML engineering and scientific discovery through agentic tree
search.

This search leaves behind far more than a final answer. Each step records
the prompt the model saw, the artifact it produced, the tool outputs it
observed, an objective reward, and causal links to its parent and
siblings---together forming what we call an \emph{experience graph}. Yet
every one of the systems above keeps this graph in Python objects and JSON
checkpoints. We learned the cost of this firsthand. KernelEvolve's first
implementation stored its search tree in process-local state and file
checkpoints; this sufficed for a single session but failed as a production
foundation. Progress was lost on crashes, parallel workers needed ad hoc
coordination, prior discoveries could not be reused systematically, and
training data had to be scraped from logs after the fact. Each missing
capability was really the same missing abstraction: the experience graph
was not a database object.

A purpose-designed database for long-horizon agentic tasks dissolves these
problems. We therefore explore the system architecture design space,
leading to a new data foundation---\textsc{Trellis}---that treats the
experience graph as first-class, governed, queryable database state rather
than ephemeral logs. The core insight is that \textit{search over
experience graphs is a database access pattern}: recovery is a query
against the frontier, cross-session reuse is a vector-seeded graph
traversal, training-data extraction is a materialized view, and agent
replay is an as-of temporal query. Trellis is our attempt to
define the database architecture for agentic tasks before accidental
abstractions become permanent.

This paper makes three important contributions. First, we define the
\textbf{self-improving agentic system} as a two-loop architecture: an
inner loop of skill-driven agent sessions, and an outer loop of RSI
tree search over a persistent data substrate that renders agents
stateless and serverless.
We then show how this framework extends
from a single-agent search to multi-agent scientific societies where
agents share hypotheses, critiques, and distilled
knowledge (Section~\ref{section:lhas}). Second, we show that
declarative, procedural, and episodic memory tiers around which
production agents have converged are an incomplete, ungoverned special
case, and describe a
concrete data foundation that unifies them with the reward-bearing
experience graph through a
query layer composing graph traversal, vector similarity, and
structured filters. In doing so, we identify common access patterns---append-heavy writes mixed with multi-hop path updates, vector retrieval, and as-of
temporal reconstruction---that
could benefit from new system architectures
(Sections~\ref{section:workload}--\ref{section:training}). Third, we
introduce new database research opportunities---multi-modal query planning,
consistency for concurrent tree search, physical design for the above-described
access pattern, governed view maintenance, bi-temporal memory, and
database semantics for multi-agent institutions. We believe these are new important areas that need more investments from the data management community
(Section~\ref{section:agenda}). Throughout this paper, we ground the research on real-world
production measurements from KernelEvolve---an agentic kernel coding framework deployed in the production environment, including a controlled study
in which cross-session memory converges 10$\times$ faster at 52\% lower
token cost while revealing an exploration--anchoring tradeoff
(Section~\ref{section:measurements}).

\section{Self-Improving Agentic Systems}
\label{section:lhas}

The first generation of AI assistants, such as GitHub Copilot and
ChatGPT, was one-shot: prompt in, answer out, session over. The
current generation, such as Claude Code~\cite{claudecode2025} and
Codex~\cite{codex2025}, is agentic: the tool reads files, runs
commands, observes errors, and iterates within a single session.
\gang{These agents are no longer session-isolated. They persist declarative
facts across sessions (\texttt{CLAUDE.md}, \texttt{MEMORY.md}),
accumulate reusable skills, maintain searchable episode archives, and
run consolidation pipelines---such as Claude's dreaming
pass---that distill and prune what they have learned. But the state
these tiers capture is \emph{knowledge}: what the agent knows, what
it prefers, how it solved a class of problems before.}

\begin{figure*}[t]
\centering
\includegraphics[width=0.9\textwidth]{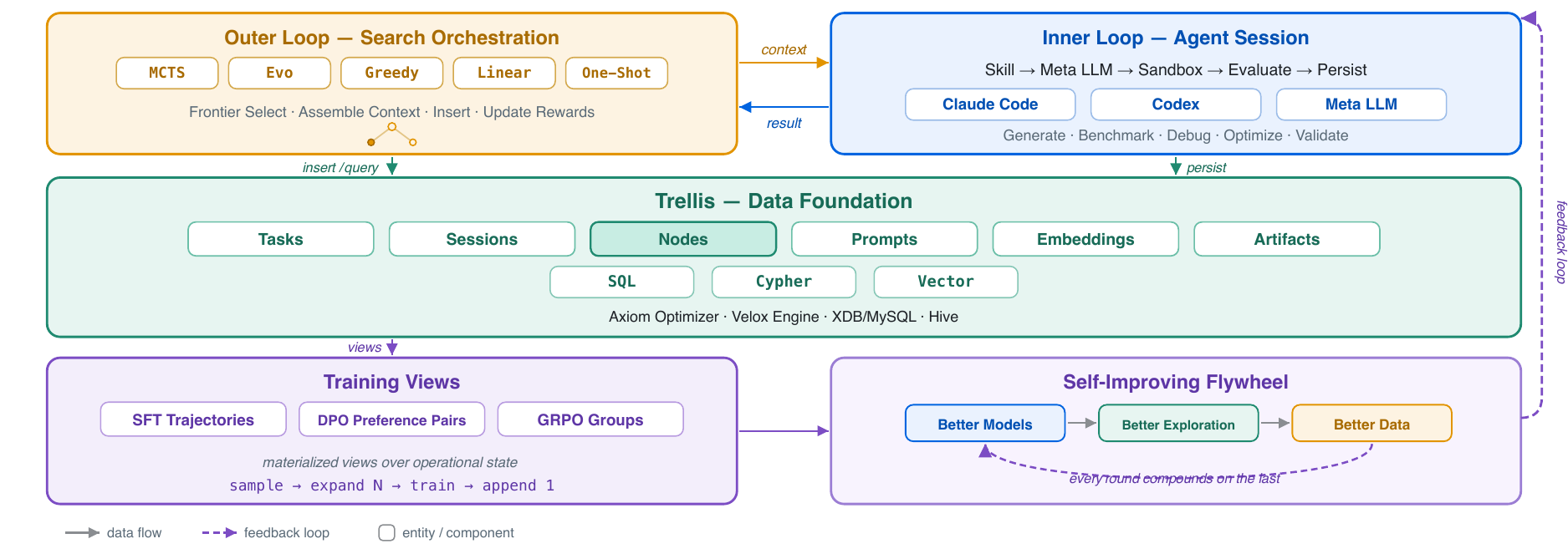}
\caption{Architecture of a self-improving agentic system.}
\label{fig:arch}
\end{figure*}

\gang{Long-horizon tasks produce a different kind of state.} Scientific
discovery, hardware design, security research, and infrastructure
optimization are problems where the right answer emerges only after
dozens or hundreds of failed attempts, each informed by the failures
before it. \gang{The state this search generates is not knowledge to be
recalled but a \emph{reward-bearing experience graph}: a causal tree
of attempts, fitness scores, sibling comparisons, and the tool outputs
that justify each decision. No file-based memory tier has a place for
this graph, and no existing agent framework exposes it as queryable,
governed, shared state.} A single session is insufficient; the system
must orchestrate \emph{many} sessions into a structured search,
accumulate results across sessions, and learn from the entire history.
This is the RSI process introduced above. We call the system that
supports RSI a \textit{self-improving agentic system}, and argue that
it has three essential components (Figure~\ref{fig:arch}): an inner
loop that executes individual steps, an outer loop that orchestrates
search, and a persistent data substrate that records everything both
loops produce.

\paragraph{The inner loop.}
Each node in the experience graph is produced by a single execution of
the inner loop---an autonomous agent session (Claude Code, Codex, or any
tool-using LLM) driven by
\textbf{skills}\footnote{Skills are reusable, user-defined task
specifications for agent sessions; see
\url{https://platform.claude.com/docs/en/agents-and-tools/agent-skills/overview}.}.
A skill tells the agent what to do, how to evaluate the result, and what
constitutes success. For example, one skill generates and benchmarks GPU kernels;
another runs timing analysis on a chip design; another measures binary
size after a compiler-pass reordering; another searches for security
vulnerabilities. The inner loop does not change across skills. It always
follows the same pattern: consume a task description and context from
the outer loop, generate candidate artifacts via LLM inference, execute
and evaluate in an isolated sandbox, and persist the structured
output---execution logs, evaluation metrics, correctness results,
analysis report, fitness score---as the node's record. The agent then
terminates; it holds no state.

\begin{figure*}[t]
\centering
\includegraphics[width=0.9\textwidth]{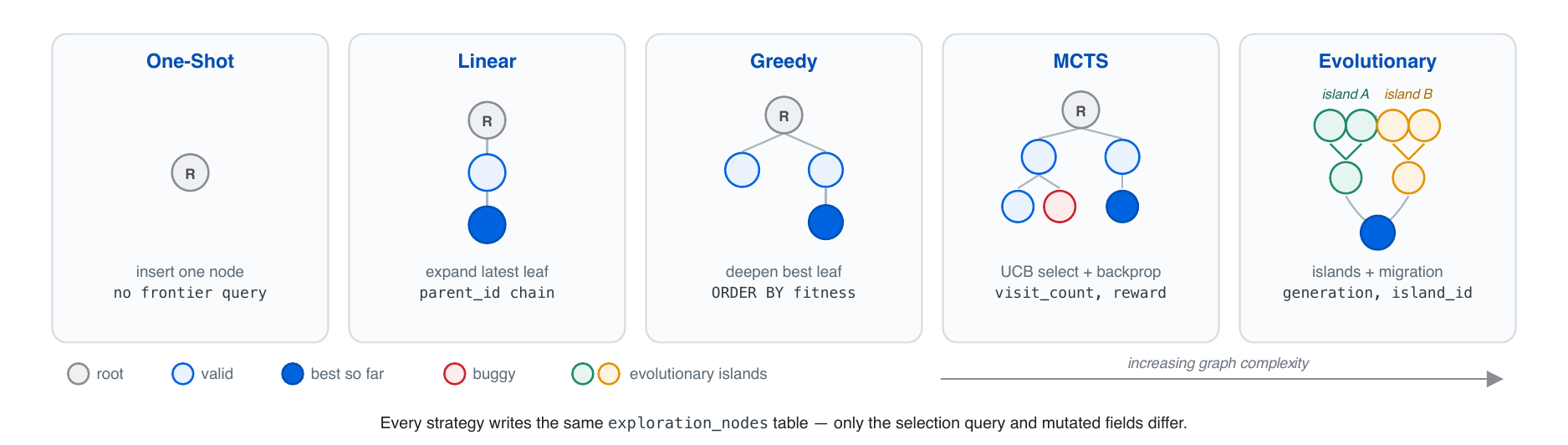}
\caption{Example search strategies and the experience graphs they produce.}
\label{fig:strategies}
\end{figure*}

\paragraph{The outer loop.}
The outer loop decides which step to take next based on a search strategy. Each step becomes a point in an experience graph. Figure~\ref{fig:strategies} shows some example search strategies and the experience graphs that they produce. A one-shot search strategy produces a single node graph. Linear strategies
chain revisions into a path. Greedy strategies draft multiple candidates, pick
the best, and deepen. MCTS strategies balance exploration and exploitation via UCB
scores~\cite{kocsis2006mcts,browne2012mcts,silver2017alphago}.
Evolutionary search maintains multiple islands with mutation, selection,
crossover, and migration, producing a
DAG~\cite{romeraparedes2024funsearch}.

The outer loop selects a frontier
node based on the search strategy, assembles context from the existing experience graph---ancestors, siblings,
prior failures---invokes the inner loop, and records the result.
It is task-agnostic:
it reads fitness scores and selects frontiers over the same data
interface, regardless of the skill.

\gang{\paragraph{The outer loop is a control plane.}
These strategies are not different systems but different policies over the
same substrate: one-shot, linear, greedy, MCTS, and evolutionary search all
read the current graph and decide which node to expand next, differing only in
how that decision is made. It is therefore a mistake to equate the outer loop
with any single algorithm. The outer loop is a \emph{control plane} that
selects a policy, and the policies form a spectrum. At one end, fixing the
outer loop to a single expansion and letting the inner loop iterate internally
yields the ``auto research'' configuration~\cite{karpathy2026autoresearch}, in
which one self-managing agent session runs the entire search and its experiment
history survives only as in-context state and a flat log---queryable by no one.
At the other, many workers expand a wide tree or
a population of islands concurrently, and collaboration across heterogeneous
harnesses (Codex, Claude Code, and others) is just another point on the same
spectrum. What is invariant across all of them is the substrate: every policy
reads frontier nodes and writes results through one data interface. The control
plane decides \emph{how} to search; the data foundation is indifferent to that
choice.}

\paragraph{Memory scaling through external state.}
The two loops produce a graph. The question is where it should live.
The central claim of this paper is that
 \emph{all} exploration state should be placed inside a persistent, queryable
store. When the database owns the graph, the system acquires memory
scaling~\cite{databricks2026memory}: performance improves as the store
grows, because new sessions query prior solutions, borrow high-fitness
subtrees, and skip dead ends. Knowledge compounds rather than
evaporates.

\paragraph{Separating compute from state.}
Persisting the graph in this way is not a mere convenience; it is the same
architectural move that has frequently reshaped data systems.
Shared-disk parallelism, cloud storage disaggregation~\cite{dageville2016snowflake}, and
serverless functions~\cite{hellerstein2019serverless,liao2023flock} all rest on one principle: once durable state lives
in a storage tier and the compute tier holds nothing it cannot afford to
lose, compute becomes elastic, interchangeable, and fault-tolerant. A
self-improving agent should be built the same way. Because every node,
reward, and frontier pointer lives in the database, the agent process
holds no irreplaceable state---it is disposable compute. Any worker can
claim any frontier node at any time; a worker that dies mid-expansion
loses at most one inner-loop invocation, never the search tree. There is
no checkpoint to deserialize, no session pinned to a machine, and no warm-up.
\gang{How much a crash costs depends on the granularity at which the graph is
externalized: a single step when every step is a node, but an entire task when
the outer loop is one and the inner loop runs long. The foundation collapses
this difference by treating the inner-loop session itself as resumable state---each
node references the session that produced it---so an interrupted session is
reattached rather than restarted. Externalization granularity thus becomes a
tuning knob rather than a correctness boundary, and the recovery guarantee
holds at either extreme.}
Recovery is a query against the frontier, and scaling is adding or
removing workers against a shared store.

\paragraph{The self-improving loop.}
The two loops, connected through a shared store, form a recursive
intelligence system. Exploration populates the store with structured
trajectories. These trajectories are materialized as training
views---supervised examples, preference pairs, reinforcement
groups~\cite{rafailov2023dpo,shao2024deepseekmath,guo2025deepseekr1}---and
used to post-train specialized models. These post-trained models drive better
exploration, which produces higher-quality data, which trains better
models. The system improves by using itself. Taken further, the data substrate
can support not just parallel workers over a single tree but a
\textbf{society} of agents that propose hypotheses, critique each other's
results, and distill lessons into shared memory---scientific
compounding, not just parallel search. None of this is possible if the
data substrate is ephemeral. Without a durable, queryable, governed
foundation, the loop cannot close.

\gang{%
\begin{table*}[t]
    \centering
    \small
    \caption{The structural differences between the experience graph and
    episodic memory}
    \vspace{-3mm}
    \begin{tabular}{@{}p{0.09\textwidth}p{0.23\textwidth}p{0.58\textwidth}@{}}
    \toprule
    & \textbf{Episodic memory} & \textbf{Experience graph} \\
    \midrule
    Records     & what the agent \emph{said} (transcript)
                & what search \emph{tried} and how well it worked (executable artifacts + objective rewards) \\
    Structure   & flat text chunks
                & tree / DAG: nodes with parent links, rewards, and siblings \\
    Retrieval   & FTS5 keyword / vector similarity
                & graph traversal (ancestor, subtree, sibling paths) + vector + structured filters \\
    Mutability  & append-only
                & mutable: \texttt{visit\_count}, \texttt{ucb\_score}, \texttt{island\_id} rewritten by search algorithms \\
    Time travel & none
                & CDC changelog $\rightarrow$ AS-OF reconstruction at any step \\
    Training    & not extracted
                & SFT trajectories, DPO sibling pairs, GRPO groups as materialized views \\
    Sharing     & per-agent, per-machine
                & governed cloud store, cross-user, cross-session \\
    \bottomrule
    \end{tabular}
    \label{table:tier34}
\end{table*}%
}

\begin{table*}[t]
    \centering
    \small
    \caption{Agent operations as queries over the data foundation, and the
    properties and requirements each relies on:
    \textbf{C}ausality, \textbf{E}xecutability, \textbf{R}eward (data
    properties); \textbf{Co}llectivity, \textbf{G}overnance (substrate
    requirements).}
    \vspace{-3mm}
    \begin{tabular}{@{}p{0.10\textwidth}p{0.73\textwidth}p{0.10\textwidth}@{}}
    \toprule
    \textbf{Operation} & \textbf{What the database must support} & \textbf{Draws on} \\
    \midrule
    Resume & Look up a session's frontier, pending jobs, and best node. & C, R \\
    Reuse & Find similar prior tasks via vector search, then retrieve their high-fitness subtrees via graph traversal. & C, R, Co \\
    Repair & Query all prior failures sharing the same error signature to avoid repeating known dead ends. & E, R, Co \\
    Train & Extract path trajectories, sibling preference pairs, and reward groups as materialized views. & C, E, R \\
    Replay & Reconstruct a node's state \emph{as of} any past step. & C, E \\
    Observe & Query search progress, reward distributions, and algorithm comparisons; render lineage graphs. & C, R \\
    Audit & Trace which prior attempts, tool outputs, and documents influenced a given decision. & C, G \\
    Govern & Enforce policy over raw traces and the artifacts derived from them---embeddings, distilled skills, and training views alike. & G, Co \\
    \bottomrule
    \end{tabular}
    \label{table:workload}
\end{table*}

\section{Why Agent Memory Is a Database Workload}
\label{section:workload}

At first blush, it may seem that agent memory can be reduced to retrieval-augmented
generation: embed text, search by nearest neighbor, and inject the results into the
context window.
Such a design would miss most of the workload.

\gang{\paragraph{Beyond files: the three-tier memory model is incomplete.}
Section~\ref{section:lhas} noted that production agents already maintain
persistent memory across sessions. To make the gap concrete, consider
the three tiers on which they have independently converged.
\emph{Declarative} memory stores durable facts and
preferences---Claude Code's
\texttt{CLAUDE.md}~\cite{claudecode2025}, Hermes's \texttt{MEMORY.md}
and \texttt{USER.md}~\cite{hermes2025}.
\emph{Procedural} memory stores reusable how-to playbooks---the
\emph{skills} of Section~\ref{section:lhas}. \emph{Episodic} memory
stores searchable records of past sessions---ranging from plain
transcripts to SQLite-backed hybrid vector-plus-FTS5 indexes with
background consolidation pipelines. Several of these systems bolt on a
process that distills and prunes the tiers---Claude Code's dreaming
pipeline, OpenClaw's three-phase consolidation, Hermes's
curator---an implicit admission that memory must \emph{evolve}, not
merely accumulate.}

\gang{These tiers are genuine engineering achievements, yet they all
capture \emph{what the agent knows}---facts, preferences, playbooks,
and past transcripts. Long-horizon search produces a fundamentally
different kind of state: a reward-bearing \textbf{experience graph}---a
causal tree of attempts, each carrying the executable artifact it
produced, the tool outputs and objective reward that scored it, the
sibling comparisons that say which variant won, and the mutable search
statistics (visit counts, UCB scores) that the algorithm rewrites as it
learns. This graph is \emph{orthogonal} to the three knowledge tiers, not
a replacement for them. The knowledge tiers tell each inner-loop agent
\emph{how} to generate a candidate; the graph records \emph{what happened
when it did}. Table~\ref{table:tier34} makes the structural gap concrete
across seven dimensions: an episodic transcript can be \emph{searched};
an experience graph can be \emph{queried, traversed, versioned, and
trained from}. Yet no file-based memory tier has a place for this graph,
and no existing agent framework exposes it as queryable, governed, shared
state.}

\dna{In a future where agents act increasingly independently, running
their own code and yielding rewards, these procedural memory stores
consisting of local files or single-embedded database are becoming hard
to manage and control centrally. This complicates querying across these
data stores.
Fundamentally, a unified, governed data foundation is needed in order to
provide an infrastructure for independent agents to work together in
harmony, all serving as workers that produce and consume a rich,
integrated dataset that includes the cross-agent experience graph.}


\gang{To see why this is a database \emph{workload}, \dna{we must consider} the \emph{properties}
of the experience-graph data, the \emph{requirements} those properties
place on any substrate data foundation that holds all four tiers together, and the \emph{operations} agents run against it.}

\smallskip\noindent\gang{\textbf{Data properties (of the experience
graph).} Three properties are intrinsic to RSI traces and absent from both
retrieved documents and the file-based tiers.}

\textbf{Causality.} A failed action only has value given the context of its parent plan,
the tool output it reacted to, and the sibling that succeeded---a path
query, not a similarity search.

\textbf{Executability.} Memories are programs, configurations, test
cases, and tool outputs---artifacts that must be replayed, diffed, and
reused, not passages to be embedded.

\textbf{Reward.} Every node carries objective feedback: correctness,
latency, throughput. Memory is an experience
buffer~\cite{lin1992experience,mnih2015dqn} for both future search and
model training---not only context for the next prompt.

\smallskip\noindent\gang{\textbf{Substrate requirements (across all four
tiers).}}

\textbf{Collectivity.} One engineer's discovery should seed every future
session on a similar workload---not merely shared high-fitness starting
points, but shared claims, evidence, critiques, and replications that
compound scientific knowledge rather than accumulate isolated data
points. \gang{This applies equally to declarative facts (a distilled rule
should be visible to every agent), procedural skills (a proven playbook
should be reusable organization-wide), episodic history (prior failures
should be queryable across users), and the experience graph (prior
subtrees should seed future search).}

\textbf{Governance.} Traces contain generated code, production metadata,
and hardware-specific signals. Policies must apply to raw traces
\emph{and} derived artifacts---embeddings, distilled rules, and training
examples. \gang{Governance is a cross-cutting constraint on the entire
foundation: a distilled skill inherits the access policy of the
episodes it was derived from, and a training view inherits the policy of
the exploration nodes it materializes.}

\smallskip\noindent\gang{\textbf{Operations.} Agents run eight operations
against this foundation (Table~\ref{table:workload}). Most are queries over
the experience graph; Reuse and Repair also reach across tiers, since
vector search spans episodic and declarative memory. The table maps each
operation to the properties and requirements it relies on.
Every one of these operations is a query over the same governed data
foundation---there is no separate search engine, training pipeline, or
observability stack.}

The resulting access pattern is unlike any existing workload.
Writes are a mix of appends (new nodes) and localized path updates (MCTS
backpropagation walks the ancestor chain to update visit counts and
cumulative rewards). Reads span four modalities: ordered scans for
frontier selection, multi-hop graph traversal for ancestor and subtree
context, vector similarity for cross-session reuse, and full table scans
for training-data extraction. No existing system---OLTP, OLAP, graph
database, or vector store---is designed for this combination. A vector
store cannot answer ``which sibling approach succeeded where this one
failed.'' A JSON checkpoint cannot support ten concurrent workers
exploring the same tree.  

\section{The Data Foundation}
\label{section:df}

No existing system is optimized to serve as the data foundation for
self-improving agents. The workload demands an operational store for
online search with sub-50\,ms latency, a graph substrate for multi-hop
traversal, and a training-data warehouse for offline learning---all
over the same governed state. The architecture therefore separates a
unified logical model from heterogeneous physical backends.

\paragraph{Logical model.}
The experience graph and episodic history---two of the four memory tiers of
Section~\ref{section:workload}---share a relational schema with a
four-level hierarchy.
\textbf{Tasks} define the problem: specification, target environment,
and success metric. \textbf{Sessions} record who is searching, with
which algorithm, and how far they have gotten. \textbf{Nodes} capture
every individual attempt: parent link, generated artifacts, execution
output, fitness score, evaluation evidence, and algorithm-specific
metadata (UCB score for MCTS, generation and island for evolutionary
search). \textbf{Prompt histories} preserve the exact messages the LLM
saw and produced. Task-description
embeddings support vector similarity search across sessions---given a
new task, the system can use these to find prior tasks with similar specifications and
retrieve their best results. Large artifacts---logs, traces,
binaries---live in object storage, linked by reference, keeping the
relational tables lean while preserving full lineage. This four-table schema is thus
algorithm-agnostic, task-agnostic, and skill-agnostic. 

\gang{\paragraph{Context as managed state.}
A recurring failure mode in tree search is context discontinuity: a new node
re-derives what its ancestors already established, wasting tokens and
reintroducing bugs that were already solved. Existing frameworks mitigate this
by threading a session identifier from parent to child, or by resuming the
parent's session in place---but there the session lives in agent memory and is
lost when the process exits. Trellis makes context a property of the substrate
instead. The \texttt{prompt\_history} table serves two roles: it
supplies the transcript that trajectory replay requires for training,
and it lets a child node either inherit its parent's session as a cheap
resume or reconstruct context from the graph through an ancestor query.
Which of the two is used is a decision for the data foundation, not a
constant wired into the agent. The context that justifies a node is thus durable, queryable, and
shared, rather than trapped in a live process.}

\paragraph{Physical architecture: separating storage from compute.}
The architecture disaggregates a stateless query and execution engine
from durable storage---the same separation that makes the agents
stateless (Section~\ref{section:lhas}), applied one level down. \dna{Although this architecture allows for any engine to be used, the current implementation uses} \gang{
Axiom~\cite{axiom2025}, a cost-based optimizer over
Velox~\cite{pedreira2022velox}, Meta's open-source vectorized execution engine.
Axiom plans SQL, Cypher~\cite{francis2018cypher}, and vector retrieval
into a single physical plan and routes each fragment to the backend
that can serve it---an operational store for sub-50\,ms frontier
queries, a vector index for similarity search, and a columnar
warehouse for training extraction---all over one logical schema.
}
Cypher exposes virtual parent--child edges over the foreign key,
enabling traversal without a separate edge table---an approach proven
effective for tree-structured metadata in distributed file
systems~\cite{liao2023filescale}---and compiles to the same engine
that serves SQL, so graph and relational queries are run over the same copy of the
data.

\paragraph{Skills and memory as evolvable shared state.}
The procedural and declarative tiers (Section~\ref{section:workload})
\dna{need to support fast reads and writes since} they are
consulted on nearly every inner-loop invocation, yet must change as
agents learn. Trellis stores them as versioned artifacts in a distributed file system, mounted read-mostly into each stateless worker over FUSE. A
worker sees an ordinary local path, while a write, such as a refined skill or a
corrected fact, commits centrally.
The background distillation that today rewrites a
machine-local \texttt{CLAUDE.md} or \texttt{MEMORY.md} thus becomes a
governed, audited update to shared state, with provenance linking each
distilled skill back to the episodes and exploration nodes that
justified it.

\paragraph{Why a unified query layer matters.}
Today's agent memory retrieval stitches together three or four service
calls at every step: a vector retrieval RPC, a structured-filter query,
a graph traversal for context, and a scope check for governance.
A graph query layer with hybrid retrieval collapses these into a single
optimizable statement---one round-trip instead of four, with the planner
choosing join order and predicate pushdown across modalities.

\dna{This is particularly important for the tree operations at the core of search---ancestor chains, subtree expansion, and sibling comparison. Walking parent pointers in application code issues $O(D)$ round-trips per lookup. Although recursive SQL operators can solve the round-trip problem, they tend to materialize every intermediate path rather than expand a frontier, and, do not currently compose with vector similarity. The Trellis graph layer instead offers first-class variable-length traversal (\texttt{[:HAS\_CHILD*1..k]}) and path predicates, and, most importantly, fuses traversal, relational filters, and vector seeding
into one planned statement.}

\paragraph{Collective intelligence via vector-seeded graph retrieval.}
Vector indexes over task-description embeddings enable cross-session
and cross-user knowledge reuse through a query pattern we call
\textbf{vector-seeded graph expansion}. Given a new task, the system
first retrieves semantically similar prior tasks via approximate
nearest-neighbor search, then follows the relational links from
matching tasks to their sessions, and from sessions to their
highest-fitness exploration nodes. A graph traversal then expands each
top-$k$ candidate into its full optimization trajectory. This pattern
composes three query modalities---vector similarity, relational joins,
and variable-length graph traversal---in a single logical request:

\begin{small}
\begin{verbatim}
MATCH (t:tasks)
WHERE t.embedding <~> $q > 0.8
MATCH (t)<-[:BELONGS_TO]-(s:sessions)
MATCH (s)<-[:IN_SESSION]-(n:nodes)
WHERE n.is_buggy = false
  AND policy_allows($user, n)
RETURN t.task_id, score(t), n.node_id,
       n.fitness_score
ORDER BY score(t) DESC, n.fitness_score DESC
LIMIT 10
\end{verbatim}
\end{small}

\noindent
This allows early sessions to
perform the hard exploration, while later sessions can begin with the best prior
solutions surfaced as reference context and refine from there. The same
retrieval also runs \emph{during} a session, injecting similar prior
nodes step by step at a tunable rate. Thus, the more sessions the store
accumulates, the less work each new one requires---a network effect at
the database level. How aggressively to inject prior nodes trades off convergence against exploration, a knob we quantify in
Section~\ref{section:measurements}.

\paragraph{Concurrent multi-agent search.}
Managing the tree in the database system allows multiple agents to explore the same
session in parallel.
Workers read frontier nodes, invoke the inner loop independently, and
insert results.

An interesting database question is what happens during
backpropagation. After evaluating a node, MCTS walks the ancestor chain
and updates each ancestor's visit count and cumulative reward. When two
workers backpropagate through a shared ancestor simultaneously, the
visit-count increment must be atomic. Stale UCB scores may cause
suboptimal frontier selection but do not compromise correctness. In
practice, the workload tolerates eventual consistency on statistics while
requiring durability on node insertion---a mixed isolation profile that
does not map cleanly to standard OLTP levels. Understanding the
right consistency semantics for concurrent tree search is an open
research question.

\paragraph{Multi-version state and time travel.}
Nodes are inserted once and never deleted, but some fields are
mutated in place: backpropagation rewrites visit counts and rewards
along the ancestor chain, and evolution reassigns a node's island.
To avoid discarding history, Trellis
captures every field-level mutation in a lightweight change log keyed by a
logical \emph{step number}, the evaluation order rather than wall-clock
time, so that progress is comparable across machines and runs.
This turns the experience
graph into a multi-version object: any past state is reconstructible by
replaying the log backward to a target step.

\gang{Multi-version state is not merely an observability convenience---it
is a correctness requirement for training. The statistics an agent acted
on---the visit counts and UCB scores that drove a frontier choice---are
exactly the fields backpropagation later overwrites; reconstructing a
trajectory from final state would train the model on decisions justified
by information that did not yet exist.}

\section{From Memory to Training}
\label{section:training}

Many agent systems bolt on training-data collection as an afterthought,
scraping logs into a separate pipeline. That approach loses structure,
provenance, and governance. In our architecture, training data is not
collected---it is \textbf{queried}. The same node table that drives online
search yields training data as materialized views.

\paragraph{SFT trajectories.}
Root-to-leaf paths through non-buggy, high-fitness nodes are multi-turn
agentic trajectories. The \texttt{prompt\_history} table preserves the exact
messages the LLM saw at each step. But message fidelity is not enough:
the numeric state that drove each decision---visit counts, UCB scores,
island membership---is mutated in place by later search, so replaying it
from final state leaks future information into the example. A faithful
trajectory reads each node \emph{as of} its own step, an AS-OF
reconstruction over the change log of Section~\ref{section:df}. A
recursive ancestor query then reconstructs any trajectory with full
provenance.

\paragraph{DPO preference pairs~\cite{rafailov2023dpo}.}
Siblings with the same parent where one has substantially higher fitness are
natural preference pairs---a graph pattern match, not a post-hoc log scrape:

\begin{small}
\begin{verbatim}
MATCH (p)-[:HAS_CHILD]->(a),
      (p)-[:HAS_CHILD]->(b)
WHERE a.fitness_score > b.fitness_score + $m
  AND a.is_buggy = false
RETURN a AS chosen, b AS rejected
\end{verbatim}
\end{small}

\paragraph{GRPO: database as episode buffer~\cite{shao2024deepseekmath}.}
Rolling out a complete search tree can take hours to days; generating
multiple trees from the same starting point for group-relative training
is prohibitively slow. The database eliminates this bottleneck. We sample a state from the persistent buffer,
generate $N$ child candidates (a GRPO group), evaluate them in the target environment,
compute group-normalized advantages, train using the policy gradient, and
append one canonical node back. Multi-turn rollout cost collapses to
single-step expansions. Advantages are computed via SQL window functions
over same-depth or same-generation cohorts.

\paragraph{Value models over experience-graph states.}
The same store defines the state space for \emph{learned control} of the
outer loop. A value model can estimate the expected gain from expanding a
given frontier node, reusing a prior subtree, merging two branches, or
invoking a verifier---conditioning not on a token sequence but on graph
features the database already holds: ancestor rewards, sibling diversity,
failure signatures, artifact diffs, retrieval scores, visit counts, and
remaining budget. The experience graph is therefore not only a replay
buffer for policy training; it is the structured state over which future
inference-time search policies are themselves learned.

\paragraph{The closed loop.}
Production exploration populates memory. Memory improves retrieval and
frontier selection. The resulting traces train future models via
SFT/DPO/GRPO~\cite{rafailov2023dpo,shao2024deepseekmath,guo2025deepseekr1}.
Improved models produce higher-quality attempts. The flywheel compounds:
more users $\rightarrow$ more states $\rightarrow$ better RL $\rightarrow$
better models $\rightarrow$ better exploration $\rightarrow$ higher-quality
states. Every production session is simultaneously a serving operation and
a training-data generation pipeline---because both are views over the same
governed store. This coupling raises a governance question: when a source
node is retracted or invalidated, the training examples derived from it
must be retracted too. The data foundation must propagate these
retractions through materialized views while respecting access policies
that may differ between the source trace and its derivatives.

\section{Early Measurements}
\label{section:measurements}

KernelEvolve~\cite{kernelevolve2026,kernelevolve2026arxiv,kernelevolve2026blog},
built on Trellis, optimizes accelerator kernels across four
hardware platforms (NVIDIA, AMD, MTIA, CPU). Its data foundation has
accumulated a substantial corpus of experience graphs spanning many tasks
and sessions. We report preliminary results across three dimensions; a
fuller evaluation is left to future work.

\paragraph{Recovery and statelessness.}
In the prior in-memory architecture, a crashed session lost all progress.
With database-native state, any worker resumes from the last committed
node: recovery is a frontier query rather than checkpoint
deserialization, with no warm-up and nothing pinned to the failed
machine. In production, sessions that hit infrastructure failures resume
automatically on another worker with no lost nodes---durability we never
engineered as a feature but inherited from putting the state in the
database.

\begin{figure}[t]
\centering
\includegraphics[width=\linewidth]{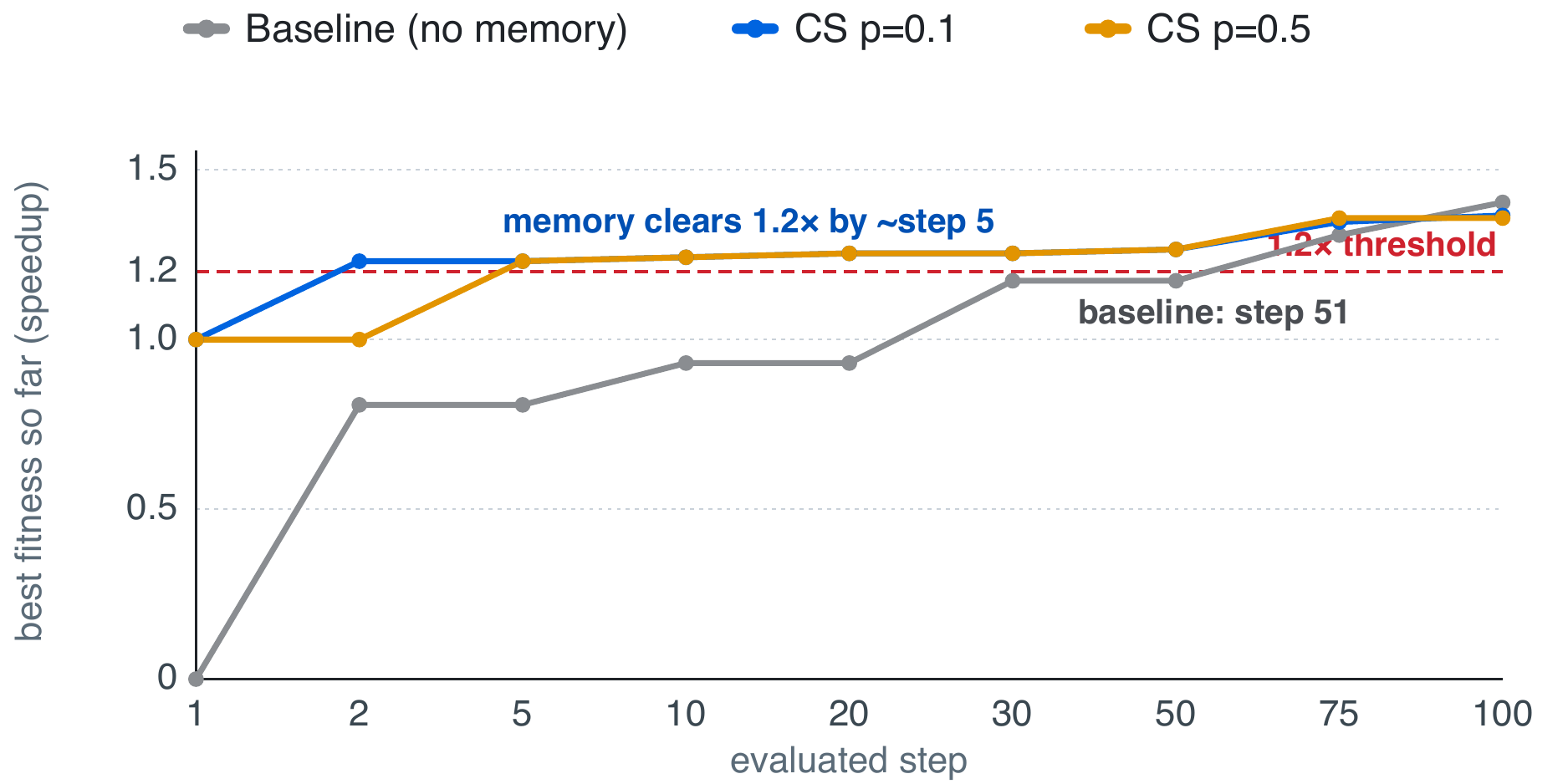}
\caption{Best fitness over evaluated steps. Both memory settings clear
$1.2\times$ within $\sim$5 steps; the no-memory baseline takes 51
steps---a $10\times$ early-convergence improvement. Memory converges
fast but plateaus early; the baseline keeps climbing and eventually
reaches the single best point ($1.49\times$), illustrating the
exploration--anchoring tradeoff that the injection policy must manage
(Section~\ref{section:agenda}).}
\label{fig:convergence}
\end{figure}

\begin{figure}[t]
\centering
\includegraphics[width=\linewidth]{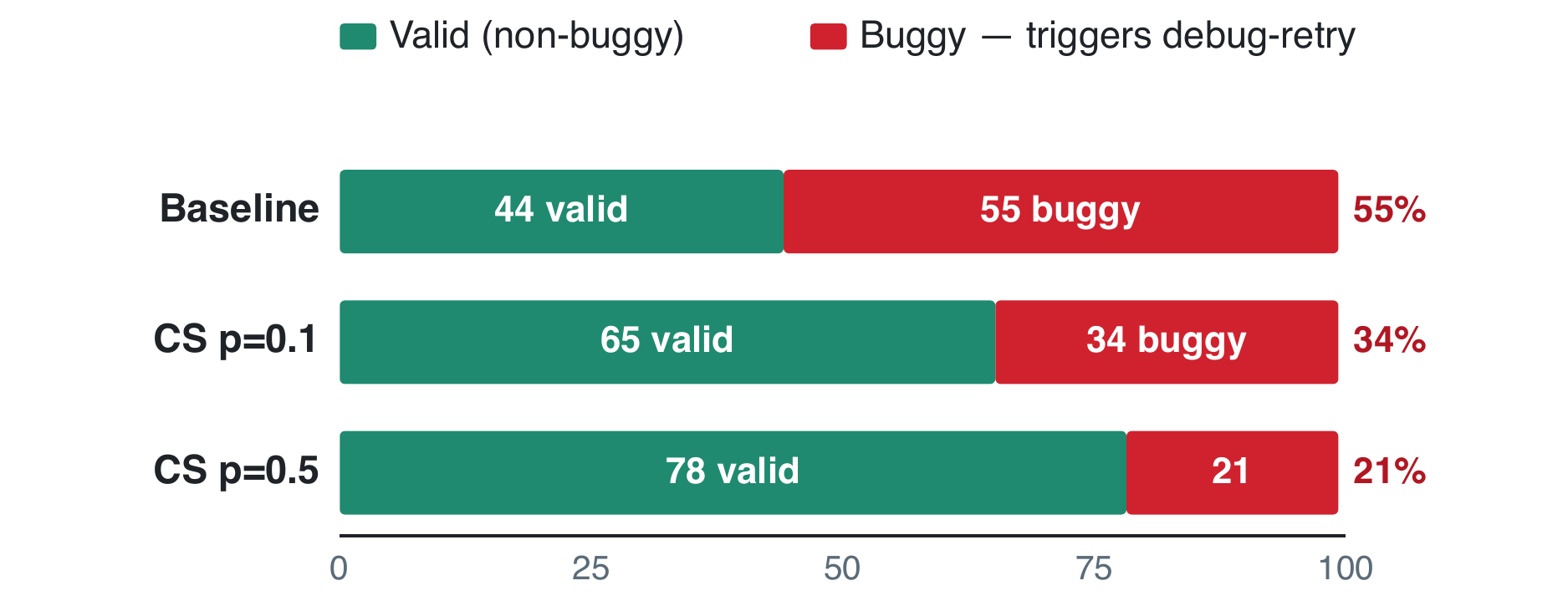}
\caption{Valid versus buggy nodes per configuration (of $\sim$100 each).
The baseline wastes 55\% of compute on buggy nodes; cross-session memory
cuts that to 34\% ($p{=}0.1$) and 21\% ($p{=}0.5$). Because each buggy
node triggers an expensive debug-retry loop, this failure reduction drives
the 52\% token-cost saving.}
\label{fig:valid-buggy}
\end{figure}

\paragraph{Cross-session reuse.}
Cross-session memory embeds each task description and node analysis,
retrieves semantically similar prior tasks and nodes, and injects them as
reference context during exploration; the injection rate $p$ is a tunable
knob. Holding the model, step budget (100 steps), worker count, and
greedy search strategy fixed, we compared no memory against $p{=}0.1$ and
$p{=}0.5$ over $\sim$100-node sessions
(Figures~\ref{fig:convergence}--\ref{fig:valid-buggy}). \gang{All figures are
averaged over three independent sessions per configuration to control for
the high variance of LLM sampling, and the trends reported below were
consistent across runs.}
Memory cuts the buggy-node rate
from 55\% to 34\% ($p{=}0.1$) and 21\% ($p{=}0.5$), raises the fraction
of valid nodes meeting the baseline speedup from 79.5\% to 90.8\% and
100\%, and reaches a 1.2$\times$ speedup within $\sim$5 steps versus 51
for a cold start---a 10$\times$ acceleration. Because buggy nodes trigger
expensive debug-retry loops, avoiding them cuts tokens per valid node by
52\%, even though each individual step costs
the same.

These gains come with a tradeoff the architecture must expose rather than
hide. Higher injection makes search more consistent (lower variance,
fewer bugs) but more anchored: at $p{=}0.5$ the agent collapses onto 8
strategy combinations versus 20 for a cold start, and the single best
solution comes from \emph{no} memory (1.49$\times$ versus 1.36$\times$).
The balanced setting $p{=}0.1$ retains most of the convergence win while
producing the most high-quality ($\geq$1.3$\times$) solutions. Memory
accelerates and stabilizes search, but unbounded reuse suppresses the
diversity that finds the best answers---making the injection policy a
first-class tuning question for any memory-backed agent
(Section~\ref{section:agenda}).

\paragraph{Training data without re-rollout.}
Every production session leaves behind complete multi-turn trajectories,
so training data---SFT paths, DPO sibling pairs, GRPO cohorts---is
extracted as queries over the same store rather than produced by a
separate pipeline (Section~\ref{section:training}). The payoff is
wall-clock. The multi-turn rollout that dominates agentic RL---where a
single search tree takes hours to days and group-relative methods need
many from the same state---collapses into reuse of trajectories the store
already holds, plus single-step expansions from a sampled state when
fresh samples are needed. Every production run is thus simultaneously a
serving operation and a training-data generator, with full provenance
(task, hardware, session, algorithm) attached.

\section{Beyond Kernels and Research Agenda}
\label{section:agenda}

KernelEvolve is one instance of a general pattern, and we have already
exercised that generality. We retargeted the same Trellis foundation
from optimizing machine learning kernels to \textit{validating MTIA silicon hardware}---a
different application use case entirely---by changing only the fitness function and
the skills. Instead of rewarding speedup, the agent is incentivized for
\emph{finding bugs}---a mismatch between the silicon simulator and the
golden reference scores high while a clean pass scores zero. The skills now
describe the instruction set architecture (ISA), known errata, and coverage-guided bug-hunting
strategies; cross-session memory tracks an ISA coverage matrix so each
session targets the gaps left by prior ones, and self-evolving search
expands a single bug into a family across dtypes, shapes, and hardware
units. The experience graph, query layer, search orchestration,
training views, and observability carried over unchanged---the
infrastructure is domain-agnostic, and the domain is just a skill. We have
exercised this silicon-validation configuration end to end. Drug
discovery, chip floorplanning, compiler optimization, and automated
scientific research~\cite{chen2021codex,park2023generativeagents} share
the same structure---tree-structured exploration traces with objective
rewards.

\paragraph{Query planning for multi-modal composition.}
Vector-seeded graph expansion (Section~\ref{section:df}) composes
vector similarity, relational joins, and variable-length graph
traversal in a single logical request. No existing query planner has a
cost model for this composition: the vector index's selectivity
determines the join fan-out, which determines the traversal cost, but
no system maintains cross-modal
statistics~\cite{johnson2019faiss,angles2017pgm}. Building a planner
that reasons across these three stages is an open problem.

\paragraph{Consistency for concurrent tree search.}
As discussed in Section~\ref{section:df}, MCTS backpropagation is a
multi-hop read-then-write that tolerates eventual consistency on
statistics but requires durability on node insertion. Formalizing the
right isolation level for this mixed profile---and proving that
relaxed consistency does not degrade search quality beyond a bounded
factor---is a question the database community is well-equipped to
answer.

\paragraph{Physical design.}
The workload combines appends, path updates, ordered reads, multi-hop
traversals, vector similarity, and full scans. No existing physical
design---row store, column store, LSM tree, or graph-native
storage---is optimized for this
combination~\cite{stonebraker2005onesize,stonebraker2018cstore}. What
indexing and layout strategies minimize total cost across all access
patterns?

\paragraph{Governed view maintenance.}
Training views are materialized from operational state
(Section~\ref{section:training}). When a source node is retracted, all
derived training examples must be invalidated---a view-maintenance
problem under governance constraints where access policies may differ
between the source trace and its
derivatives~\cite{pedreira2022velox,francis2018cypher}.

\paragraph{Retrieval policy and memory quality.}
Section~\ref{section:measurements} showed that how much prior memory to
inject is a tunable knob with a sharp tradeoff: more reuse converges
faster and fails less, but anchors the search and suppresses the
diversity that finds the best solutions. This is a retrieval-planning and
data-quality problem in disguise. What is the right injection
policy---fixed, decayed over a session, or chosen per query from the
selectivity and confidence of the match? How should the store score
memory quality, detect stale or low-quality entries that would reinforce
past mistakes~\cite{databricks2026memory}, and keep retrieval fresh as
hardware and workloads drift? Treating reuse as a cost-based,
quality-aware retrieval decision rather than a fixed heuristic is an open
problem the database community is well-positioned to tackle.

\paragraph{Bi-temporal memory.}
The change log of Section~\ref{section:df} versions exploration state
along a single axis---evaluation order---which already supports time
travel over a search. Full bi-temporal
semantics add a second axis, separating \emph{valid time} (when a fact
was true) from \emph{transaction time} (when the entry was committed),
unlocking late-arriving corrections, memory-drift detection across
environment changes, and distillation audit (``what did the agent know
at time $T$, and was it still true?''). Lifting versioning from a
field-level log to first-class temporal predicates in the query
layer---and planning queries that range over both axes---is an open
problem.

\paragraph{Learned control over experience graphs.}
A database of experience graphs turns the outer loop from a fixed search
heuristic into a policy learned from accumulated experience. Value models
over graph state (Section~\ref{section:training}) can predict which
frontier node, retrieved subtree, verifier call, or repair is worth the
next unit of compute, making the database the state representation for
inference-time control. What graph features, freshness guarantees, and
training objectives such a controller needs---and how to serve its
predictions within the agent loop's latency budget---are open questions.

\paragraph{From parallel search to scientific societies.}
The concurrent search model in Section~\ref{section:df} treats agents
as independent workers over a shared tree. A more ambitious design
organizes agents as a \textit{scientific society}: ideator agents
propose hypotheses, builder agents implement them, reviewer agents
critique results and request replications, and distiller agents
consolidate lessons into reusable memory. Each institution in this
society---a leaderboard that separates score from confidence, an idea
store whose claims carry state machines (proposed $\to$ tested $\to$
replicated $\to$ distilled), a structured forum for evidence-linked
peer review, and a scheduler that converts disagreement into compute
allocation---maps to database tables, state transitions, and queries.
The data foundation becomes the institutional infrastructure for
collective intelligence. Designing the transactional semantics,
consistency models, and governance policies for such multi-agent
institutions is, we believe, among the most consequential open
problems at the intersection of databases and AI.

\section{Related Work}
\label{section:related}

\gang{Trellis sits at the intersection of agent memory, recursive
self-improvement, graph/vector data management, and training-data
infrastructure.}

\gang{\textbf{Agent memory systems.}
MemGPT~\cite{packer2023memgpt} introduced the OS analogy---paging
information between fast (in-context) and slow (external)
memory---establishing that agents need a memory hierarchy, not just a
larger window. Trellis takes the complementary \emph{database}
analogy: MemGPT is a transport layer; Trellis is the storage engine
underneath---how you structure, index, query, version, and govern the
data that lives outside context. Dedicated agent-memory services such
as Mem0~\cite{chhikara2025mem0} add structured long-term memory;
Graphiti~\cite{rasmussen2025graphiti} goes further with a temporal
knowledge graph that dynamically synthesizes conversational data---the
closest prior work in spirit, though aimed at enterprise memory rather
than RSI experience graphs and without CDC, materialized training
views, or a graph-native query layer. Production coding agents have
moved further still with declarative facts, procedural skills, and
episodic transcripts (Section~\ref{section:workload}), yet these remain
per-agent local files or embedded databases with no cross-user query
and no governance. A recent comprehensive
survey~\cite{hu2025memorysurvey} calls for treating memory as a
``first-class primitive''---precisely Trellis's thesis, which the
survey motivates but does not realize as a database system.}

\gang{\textbf{Recursive self-improvement systems.}
AlphaEvolve~\cite{novikov2025alphaevolve},
FunSearch~\cite{romeraparedes2024funsearch},
AIDE~\cite{jiang2025aide}, AIRA~\cite{toledo2025aira}, the AI
Scientist~\cite{yamada2025aiscientist}, and open frameworks such as
OpenEvolve~\cite{openevolve2025} show that long-horizon exploration can
produce valuable results, building on code-generation
foundations~\cite{chen2021codex} and multi-agent
simulation~\cite{park2023generativeagents}. All keep their traces as
application state---Python objects and JSON checkpoints. Trellis
treats those traces as the primary database workload.}

\gang{\textbf{Experiment tracking and ML metadata.}
Run trackers such as MLflow~\cite{zaharia2018mlflow} and
ModelDB~\cite{vartak2016modeldb}, and dataset/metadata catalogs such as
Google Goods~\cite{halevy2016goods}, record runs, hyperparameters,
metrics, and artifact lineage, and might appear to already cover the
experience graph. They do not. These are offline, sidecar registries that
log metrics \emph{about} runs after the fact. They do not serve the
online, sub-50\,ms frontier queries that drive an agent's next step; they
do not model the mutable search statistics (visit counts, UCB scores) that
search rewrites in place; they offer no as-of reconstruction of the state
an agent acted on; and they provide no graph-native ancestor/sibling
traversal fused with vector retrieval. Trellis is the operational store
that drives search and materializes training views from the same state,
not a record of experiments past.}

\gang{\textbf{Agent orchestration.}
Meta-harness systems such as Omnigent~\cite{zaharia2026omnigent} lift
orchestration above individual harnesses, providing multi-agent
composition, policy governance, and real-time collaboration.
Sakana Fugu~\cite{sakana2026fugu} trains an orchestrator to dynamically
compose agentic scaffolds over a pool of frontier workers, with
inter-workflow shared memory and GRPO-based training---demonstrating
that multi-agent coordination increasingly requires persistent, shared
state. These systems validate the need for a shared layer above agents
but operate at the control plane---session routing, cost budgets,
cross-vendor review. The data plane---how to persist, query, version,
and train from the exploration state these agents produce---remains
unaddressed and is precisely the gap Trellis fills.}

\gang{\textbf{Graph/vector data management.}
Graph databases~\cite{angles2017pgm} and
Cypher~\cite{francis2018cypher} provide the language of paths and
neighborhoods. Vector indexes~\cite{johnson2019faiss} provide semantic
retrieval. The Trellis workload needs both at once, composed with
structured filters, transactional updates, and temporal
reconstruction. Its contribution is not a new graph language or a new
vector index, but the composition required by agent memory.}

\gang{\textbf{RL data and temporal databases.}
Experience replay~\cite{lin1992experience,mnih2015dqn},
DPO~\cite{rafailov2023dpo}, and GRPO~\cite{shao2024deepseekmath}
motivate the training views. The database contribution is provenance
and fidelity: examples are extracted from the operational store with
rewards, siblings, trajectories, and AS-OF state rather than scraped
after the fact. Temporal
databases, MVCC, and CDC provide the historical lineage for the
time-travel design. Trellis applies those old ideas to a new
object: the mutable search state of a self-improving agent.}

\section{Conclusion}

Self-improving agent memory is a database problem. Larger models and
longer context windows will not solve it. What self-improving agents need is a durable,
governed data foundation where the experience graph---every artifact,
reward, decision, and causal link---lives as queryable, shared,
training-ready state. We design, implement, and demonstrate \textsc{Trellis}
for long-horizon agentic coding tasks, running in production. The
properties that matter most---crash recovery, horizontal scaling,
cross-user reuse, and training-view materialization---are addressed by an
application-specific database design rather than engineered as one-off
features.
But the deeper prize is not infrastructure efficiency; it is collective
intelligence. Today's self-improving systems are individual agents
searching alone over private trees. The architecture we describe makes
it possible for populations of agents to share structured knowledge,
critique each other's results, and compound discoveries across sessions,
users, and time. When the data foundation becomes the institutional
infrastructure for multi-agent scientific societies---with leaderboards,
structured peer review, and governed distillation---the result is not
just agents that remember but agents that learn as a community.

The decades of research in databases have defined clean, functional interfaces and system architectures for transactions,
warehousing, streaming, and graphs that power the Internet. Agentic memory is the next wave of workloads, demanding new interfaces and database designs, thereby shaping the future of AI. Logs made databases reliable; experience graphs may make agents cumulative.

\begin{acks}
We thank Jia Jiunn Ang, Ming Chen, Barry Dong, Amit Dutta, Yuanwei Fang,
Zacharias Fisches, Vishal Gandhi, Alicia Golden, Chanoch Goldfeder,
Wei Guo, Samuel Hsia, Rohit Jain, Jacob Kahn, Michael Kuchnik, Richard Li,
Yiting Li, Jimmy Lu, Keyur Muzumdar, Michael Norris, Dmitrii Pedchenko,
Honghao Qiu, Andrii Rosa, Shiqin Wang, Ruichao Xiao,
Chao Xie, Yavuz Yetim, Hongtao Yu, and Roger Zheng for discussions and feedback
that helped shape the systems agenda in this paper.
\end{acks}

\bibliographystyle{ACM-Reference-Format}
\bibliography{paper}

\end{document}